\documentclass{article}
\usepackage{amsmath}
\usepackage{mathrsfs}
\usepackage{amssymb}
\usepackage{graphicx}
\begin{document}
\noindent
{\sl\Large Apparent Horizons with Nontrivial Topology and the Hyperhoop Conjecture 
in Six-Dimensional Space-Times
}

\vspace{12pt}
\noindent
{\sl\large Tomohiro Kurata, Hirotaka Nakayama and Takahiro Okamoto}

\vspace{6pt}
\noindent
{\sl Department of Physics, Gakushuin University, Tokyo 171-8588, Japan}

\vspace{12pt}
{\it
We investigate the validity of 
the hyperhoop conjecture,
which claims to determine a necessary and sufficient condition for the
formation of  black hole horizons in higher-dimensional space-times.
Here we consider momentarily static, 
conformally flat initial data sets 
each describing a gravitational field of uniform massive $k$-sphere sources,
for $k=1,2$,
on the five-dimensional Cauchy surface. 
The numerical result shows
the validity of the hyperhoop conjecture for a wide range of model parameters.
We also confirm for the first time the existence of
an apparent horizon homeomorphism to $S^2\times S^2$ or $S^1\times S^3$,
which is a higher-dimensional generalization of the black ring.
}

\section{Introduction}
The hoop conjecture is a well-known statement in general relativity.
This claims that
black holes with horizons form when, and
only when, a mass $M$ gets compacted into a region whose
circumference $C$ in every direction is $C \hspace{0.3em}\raisebox{0.4ex}{$<$}\hspace{-0.75em}\raisebox{-.7ex}{$ \sim $}\hspace{0.3em}
 4\pi GM$ \cite{Thorne}.
The hoop conjecture implies that a black hole horizon should not far deviate geometrically
from the round sphere. 
However, such a statement is only valid for four-dimensional space-times;
in fact, even infinitely long black string solutions 
are allowed by the Einstein equation
 in five or higher-dimensional space-times.

The higher-dimensional generalization of 
the hoop conjecture, which is called the {\it hyperhoop conjecture},
has been known for almost a decade~\cite{Ida-Nakao}, 
and its validity has been verified in various cases ~\cite{Barrabes,Yamada-Shinkai,Chul-Nakao-Ida,Yoshino-Numbu,Cvetic}.
Concretely, the hyperhoop conjecture for $(m+1)$-dimensional space-times
claims that black hole horizons form when, and only when, 
the mass $M$ gets compacted within a region 
such that 
for any characteristic $(m-2)$-dimensional closed submanifold of the region
with the volume $V_{m-2}$,
the
inequality
\begin{eqnarray}
V_{m-2} \hspace{0.3em}\raisebox{0.4ex}{$<$}\hspace{-0.75em}\raisebox{-.7ex}{$ \sim $}\hspace{0.3em} G_{m+1} M,
\label{hyper-hoop-conjecture}
\end{eqnarray}
holds,
where $G_{m+1}$ is the gravitational constant for $(m+1)$-dimensional theory of gravity~\cite{Ida-Nakao}. 
A similar condition is also investigated in Ref.~\cite{Barrabes}.

Ida and Nakao~\cite{Ida-Nakao} investigate various types of initial data sets
for five-dimensional black hole space-times, and give a physical grounds 
to believe
that
 the inequality~(\ref{hyper-hoop-conjecture}) is a good criterion for the black hole formation.
They consider gravitational fields of spatially extended objects in 
momentarily static, conformally flat initial data sets, evaluate 
various hyperhoops for apparent horizons, which they find, and give evidences 
for the validity of the hyperhoop conjecture.
Other numerical works also suggest the validity of the hyperhoop conjecture~
\cite{Yamada-Shinkai,Chul-Nakao-Ida,Yoshino-Numbu}. 

On the other hand, it is known that a higher-dimensional black hole can have
various kinds of nontrivial topologies~\cite{Cai-Galloway}, although in four-dimensional space-times
the apparent horizon must be diffeomorphic to the 2-sphere~\cite{Haw}.
The only topological restriction on apparent horizons in higher-dimensional space-times,
which is known,
is that they must admit a Riemannian metric with positive scalar curvature~\cite{Galloway-Schoen}.
It follows that a black hole in five-dimensional space-time must be diffeomorphic
to $S^3$, $S^2\times S^1$ or elliptic spaces.
In six or higher-dimensional space-times, we know little about such topological restrictions.
In five-dimensional space-time, the exact solution describing the stationary
rotating black ring, which is a black hole with $S^2\times S^1$ horizon, is
known for the vacuum Einstein equation~\cite{Emp}.
One might expect the existence of higher-dimensional generalizations of
black rings, which is a black hole with a $S^k\times S^l$ horizon,
although we do not have such an example.
Here, we try to find the ``hyper black ring'' in an initial data set
for the six-dimensional space-time, which is another motivation of the present work.

In this paper, 
we consider initial data sets for the six-dimensional black hole space-times and 
investigate the validity of the hyperhoop conjecture.

In Sec. I\hspace{-.1em}I ,  
we construct the momentarily static initial data sets on the five-dimensional Cauchy surface. 
In Sec. I\hspace{-.1em}I\hspace{-.1em}I , we 
describe the method to determine the location of the apparent horizon.
In Sec.  I\hspace{-.1em}V, we show numerical results.

The speed of light is set to unity.

\section{Hamiltonian and Momentum Constraints}

We set the initial data set $(g_{\mu\nu},K_{\mu\nu})$ on the five-dimensional Cauchy surface $\Sigma^5$,
where $g_{\mu\nu}$ is the induced metric on $\Sigma^5$ and 
$K_{\mu\nu}$ is the extrinsic curvature of  $\Sigma^5$.
Denoting by $n_\mu$ the unit normal vector to $\Sigma^5$, $K_{\mu\nu}$ is written as
\begin{eqnarray}
K_{\mu\nu}=g_\mu{}^\lambda {}^{(6)} \nabla_\lambda n_\nu,
\end{eqnarray}
where ${}^{(6)} \nabla_\mu$ is the covariant derivative associated with the
space-time metric.

The initial data set $(g_{\mu\nu},K_{\mu\nu})$ satisfies the Hamiltonian and momentum constraints  
\begin{eqnarray}
R-K_{\mu\nu} K^{\mu\nu}+K^2=16 \pi G_6 \rho  \label{Hamiltonian-constraint-1} 
\end{eqnarray}
and
\begin{eqnarray}
D^\nu (K_{\mu\nu}-K g_{\mu\nu})=8\pi G_6 J_\mu,  \label{momentum constraint-1}
\end{eqnarray}
where $R$ is the Ricci scalar associated with $g_{\mu\nu}$,  
$D^\mu$ is the covariant derivative on $\Sigma^5$, 
$G_6$ is the gravitational constant in a six-dimensional theory of gravity,
$\rho$ is the energy density
and $J_\mu$ is the 
energy flux.
Let us consider the momentarily static initial data set
\begin{eqnarray}
K_{\mu\nu}=0
\end{eqnarray}
and assume the conformally flat metric
\begin{eqnarray}
g_{\mu\nu}=\psi^{4/3} \delta_{\mu\nu}.
\end{eqnarray}
Then the energy flux $J_\mu$ is equal to zero by momentum constraint (\ref{momentum constraint-1}).
On the one hand, the Hamiltonian constraint (\ref{Hamiltonian-constraint-1}) becomes
\begin{eqnarray}
{}^{(5)}\nabla^2 \psi=-3\pi G_6 \psi^{7/3} \rho,  \label{Hamiltonian-constraint-1-1}
\end{eqnarray}
where ${}^{(5)}\nabla^2$ is the flat space Laplacian.
We solve Eq.~(\ref{Hamiltonian-constraint-1-1}) with boundary condition 
\begin{eqnarray}
\psi=1~~~~(\mbox{at infinity}),
\end{eqnarray}
and consider the vacuum case $\rho=0$,
\begin{eqnarray}
{}^{(5)}\nabla^2 \psi=0.
\end{eqnarray}

\section{The Location of The Apparent Horizon }
In this paper, we consider the uniform massive $k$-sphere source $(k=1,2)$ in Cauchy surface $\Sigma^{5}$ and investigate whether the apparent horizon $H$ which is homeomorphic with $S^{k} \times S^{4-k}$  satisfies the hyperhoop conjecture or not. For this purpose, we derive the equation which determines the location of $H$. Since we assume a momentarily static initial data set, the location is given by the minimal surface. This fact means the trace of the extrinsic curvature of $H$ is zero.\\

Because we consider $\Sigma^{5}$ is conformally flat space, we can write the metric 
$\boldsymbol{g}$ as follows:  
\begin{eqnarray}
\boldsymbol{g}=\psi^{4/3} \left[dx^{2} + x^{2}
\boldsymbol{\gamma}(S^k)
+dy^{2} + y^{2}
\boldsymbol{\gamma}(S^{3-k})
\right],
\label{spatial_metric}
\end{eqnarray}
where 
$\boldsymbol{\gamma}(S^k)$
is the metric of $k$-sphere and 
$\boldsymbol\gamma(S^{0})$ 
is regarded as zero.
Now we assume that massive $k$-sphere is located at $x=C,y=0$. 
Because of the rotational symmetry of the configuration in our setup, we have only to consider 
how a possible apparent horizon $H$ is embedded in the $(x,y)$ plane.
This embedding will be given as follows: 
\begin{eqnarray}
x=r(\xi)\cos\xi +C,~y=r(\xi)\sin\xi .
\end{eqnarray}
Then one can find that the induced metric on $H$ has the following form: 
\begin{eqnarray}
&&
\!\!\!\!\!\!\!\boldsymbol{h}\!=\!\psi^{4/3}\!\left[ \!\left\{\!(r_{, \xi})^{2}\!+\! r^{2}\!(\xi)\!\right\}\!d\xi^{2}\!
\!+\! x^{2}\!(\xi)
\boldsymbol{\gamma}(S^k)
\!+\! y^{2}\!(\xi) 
\boldsymbol{\gamma}(S^{3-k})
 \right]\!. \nonumber\\
\label{eq:induced-metric}
\end{eqnarray}
The components of the unit normal 1-form $\bar{n}=\bar{n}_xdx+\bar{n}_ydy$ to $H$ are given by
\begin{eqnarray}
&&\bar{n}_{x}=  \psi^{2/3}\frac{r_{,\xi}\sin \xi + r \cos \xi}{\sqrt{(r_{,\xi})^{2}+r^{2}}},\\
&&\bar{n}_{y}=  \psi^{2/3}\frac{r_{,\xi}\cos \xi - r \sin \xi}{\sqrt{(r_{,\xi})^{2}+r^{2}}}.
\end{eqnarray}

An apparent horizon in the momentarily static Cauchy surface is a minimal surface on it,
which is given by the minimal surface equation
\begin{eqnarray}
h^{\mu\nu}D_{\mu}\bar{n}_\nu=0.
\end{eqnarray}

In the present case, this is explicitly given by
\begin{eqnarray}
&&r_{,\xi \xi}\!-\!(5\!-\!k)\frac{(r_{,\xi})^{2}}{r}\!-\!(4\!-\!k)r\!-\!\frac{(r_{,\xi})^{2} \!+\!r^{2}}{r}  \nonumber \\
&&\hspace{0.1cm}\times \left( k\frac{r_{,\xi}\sin \xi \!+\! r\cos \xi}{r\cos \xi +C } \!-\! (3\!-\!k)\frac{r_{,\xi}}{r}\cot \xi \right)\!-\!\frac{8}{3}\frac{(r_{,\xi})^{2} \!+\!r^{2}}{r} \nonumber \\
&&\hspace{0.1cm}\times \left\{(r_{,\xi}\sin \xi \!+\! r\cos \xi )\frac{\psi_{,x}}{\psi}\!-\!(r_{,\xi}\cos \xi \!-\! r\sin \xi )\frac{\psi_{,y}}{\psi}\right\} \nonumber \\
&&=0.
\label{location-of-apparenthorizon-k}
\end{eqnarray}
Thus for given conformal factor $\psi(x,y)$, the problem to find the apparent horizon is reduced to the
2-point boundary value problem of the
second-order ordinary differential equation for $r=r(\xi)$.

Our main task is to find the apparent horizon and to confirm the
inequality
\begin{eqnarray}
  V_3(S) \lesssim G_6 M
\label{eq:hyper-hoop}
\end{eqnarray}
for every typical 3-volume $S$ on the horizon.

\section{Numerical Results}

\subsection{Spherical black holes}

For later convenience, let us consider the spherically symmetric configuration first.
The initial data metric $\boldsymbol{g}$ for a point source will be given by
\begin{eqnarray}
\boldsymbol{g}=\psi^{4/3}(r)\left[dr^2+r^2 \boldsymbol{\gamma}(S^4) \right].
\end{eqnarray}
Solving the Poisson equation~(\ref{Hamiltonian-constraint-1-1})
for the point source
\begin{eqnarray}
\psi^{7/3}\rho=\frac{3M_{\rm ADM}\delta(r)}{8\pi^2 r^4} ,
\end{eqnarray}
where $M_{\rm ADM}$ is the Arnowitt-Deser-Misner (ADM) mass, 
the conformal factor $\psi(r)$ 
is obtained as
\begin{eqnarray}
\psi=1+\frac{3G_6M_{\rm ADM}}{8\pi r^3}.
\end{eqnarray}
In fact, this is initial data for six-dimensional Schwarzschild-Tangherlini
black hole space-time.

The location of the apparent horizon  is determined by
\begin{eqnarray}
(r\psi^{2/3})_{,r}=0 .
\end{eqnarray}
This is easily solved to give the round 4-sphere 
\begin{eqnarray}
r=r_s:=\left(\frac{3G_6M_{\rm ADM}}{8\pi} \right)^{1/3} .
\end{eqnarray}

The circumference $L(C_s)$ of the great circle, 
the three-dimensional volume $A(S_s)$ of the equatorial 3-sphere
and four-dimensional volume $V(H_s)$ of the apparent horizon $H_s$
become
\begin{eqnarray}
\!\!\!\!\!\!\!\!\!\!\!\!\!\!\!\!\!L(C_s)\!\!&=&\!\!2\pi r_s\psi_s^{2/3}\!=\!2\pi\!\left(\frac{3G_6M_{\rm ADM}}{2\pi} \right)^{\!\!1/3}\!\!, \\
A(S_s)\!\!&=&\!\!2\pi^2(r_s\psi_s^{2/3})^3\!=\!3\pi G_6M_{\rm ADM}, \\
V(H_s)\!\!&=&\!\!\frac{8\pi^2}{3}(r_s\psi_s^{2/3})^4\!=\!\frac{8\pi^2}{3}\!\left(\frac{3G_6M_{\rm ADM}}{2\pi} \right)^{\!\!4/3}\!\!,
\end{eqnarray}
respectively, where $\psi_s:=\psi(r_s)$.

\subsection{$S^2\times S^2$ apparent horizons}

Here we consider the gravitational field produced by the uniform massive 2-source.
For this purpose, we take the spatial metric as
\begin{eqnarray}
\boldsymbol{g}=\psi^{4/3} \!\left[dx^{2} 
\!+\! x^{2}(d\theta^2\!+\!\sin^2\theta d\phi^2) 
\!+\!dy^{2} \!+\! y^{2}d\eta^2
\right],   
\end{eqnarray}
which corresponds to the case of $k=2$ in Eq.~(\ref{spatial_metric}).

Let us consider the uniform massive $2$-sphere source of the Euclidean radius $C$ given by
\begin{eqnarray}
\psi^{7/3}\rho=\frac{M_{\rm ADM}\delta(x-C)\delta(y)}{8\pi^2 C^2y}.
\end{eqnarray}
By solving Eq.~(\ref{Hamiltonian-constraint-1-1}),
the conformal factor is obtained in an analytic form as
\begin{eqnarray}
&&\!\!\!\!\!\!\!\psi\!=1\!+\!\frac{3G_6M_{\rm ADM}}{16\pi} \!\!\int^{\pi}_0\!\!\!d\theta\frac{\sin\theta}{[(x\!-\!C\cos\theta)^2\!+\!y^2\!+\!C^2\sin^2\theta]^{3/2}}\nonumber\\ 
&&\!\!\!\!\!\!\!\hspace{1.0em}\!=\!1\!+\!\frac{3G_6M_{\rm ADM}}{16\pi}\!\frac{1}{Cx} \!\!\left(\!\!\frac{1}{\sqrt{(x\!-\!C)^2\!+\!y^2}}\!\!-\!\!\frac{1}{\sqrt{(x\!+\!C)^2\!+\!y^2}}\!\!\right)\!\!. \nonumber\\
\end{eqnarray}

Now we try to search the location of the apparent horizon in the form
\begin{eqnarray}
  x(\xi)=r(\xi)\cos\xi+C,~y(\xi)=r(\xi)\sin\xi.  \label{eq:S2timesS2_ring_embedded}
\end{eqnarray}
Then, the minimal surface equation becomes
\begin{eqnarray}
&&r_{,\xi \xi}-3\frac{(r_{,\xi})^{2}}{r}-2r-\frac{(r_{,\xi})^{2} +r^{2}}{r}  \nonumber \\
&&\hspace{0.2cm}\times\left( 2\frac{r_{,\xi}\sin \xi \!+\! r\cos \xi}{r\cos \xi \!+\!C }\!-\!\frac{r_{,\xi}}{r}\cot \xi \right)\!-\!\frac{8}{3}\frac{(r_{,\xi})^{2} \!+\!r^{2}}{r} \nonumber \\
&&\hspace{0.2cm}\times\left\{(r_{,\xi}\sin \xi \!+\! r\cos \xi )\frac{\psi_{,x}}{\psi}\!-\!(r_{,\xi}\cos \xi \!-\! r\sin \xi )\frac{\psi_{,y}}{\psi}\right\} \nonumber\\
&&\hspace{0.1cm}=0.
\label{eq:S2-the-location-of-apparent-horizon}
\end{eqnarray}

If we find the solution to Eq.~(\ref{eq:S2-the-location-of-apparent-horizon}),
which is an embedded curve with its endpoints both on the $x$ axis,
it corresponds to a minimal surface homeomorphic to $S^2\times S^2$, but possibly
with conical singularities on the $x$ axis.
From the smoothness of the solution of the Eq.~(\ref{eq:S2-the-location-of-apparent-horizon}) on $x$ axis,
$r_{,\xi}$ must be equal to zero for $\xi=0$, $\pi$.
Then, the solution to this 2-point boundary value problem will give
the $S^2\times S^2$ apparent horizon.

We also search for the $S^4$ apparent horizon in the form
\begin{eqnarray}
  x(\xi)=r(\xi)\cos\xi, ~ y(\xi)=r(\xi)\sin\xi,  \label{eq:S2timesS2_sphere_embedded}
\end{eqnarray}
where $\xi$ ranges from $0$ to $\pi/2$.
Then we solve the minimal surface equation
\begin{eqnarray}
&&r_{,\xi \xi}-3\frac{(r_{,\xi})^{2}}{r}-2r-\frac{(r_{,\xi})^{2} +r^{2}}{r}  \nonumber \\
&&\hspace{0.2cm}\times\left( 2\frac{r_{,\xi}\sin \xi \!+\! r\cos \xi}{r\cos \xi}\!-\!\frac{r_{,\xi}}{r}\cot \xi \right)\!-\!\frac{8}{3}\frac{(r_{,\xi})^{2} \!+\!r^{2}}{r} \nonumber \\
&&\hspace{0.2cm}\times\left\{(r_{,\xi}\sin \xi \!+\! r\cos \xi )\frac{\psi_{,x}}{\psi}\!-\!(r_{,\xi}\cos \xi \!-\! r\sin \xi )\frac{\psi_{,y}}{\psi}\right\} \nonumber\\
&&\hspace{0.1cm}=0,
\end{eqnarray}
with the boundary condition $r_{,\xi}=0$ for $\xi=0,\pi/2$.

\begin{figure}[h]
\includegraphics[width=0.8\linewidth]{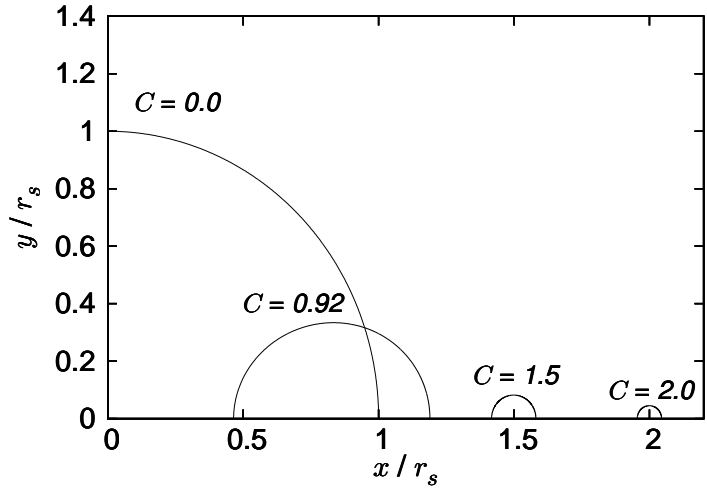}%
\caption{$S^2\times S^2$ apparent horizons for $C=0.92r_s$, $1.5r_s$, $2.0r_s$ are 
depicted in the $(x,y)$ plane.
The horizon of the spherically symmetric black hole with the same ADM mass is also depicted.
We do not find the $S^2\times S^2$ apparent horizon for $C=0.91r_s$.\label{fig:6ring-shape}}
\end{figure}

\begin{figure}[h]
\includegraphics[width=0.8\linewidth]{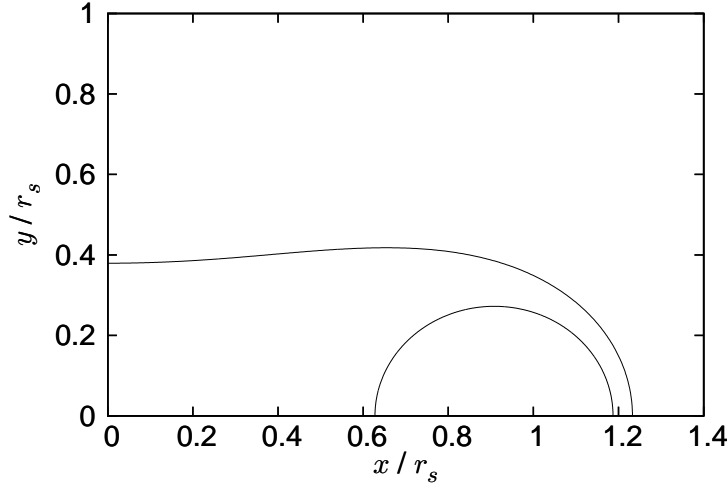}
\caption{We can find both $S^4$ and $S^2\times S^2$ apparent horizons for $0.92r_s\leq C\leq 0.96r_s$. Both apparent horizons for $C=0.95r_s$ are depicted in the $(x,y)$ plane. We cannot find an $S^4$ apparent horizon for $C=0.97r_s$.\label{fig:6ring-and-sphere}}
\end{figure}

We show some numerical results in the Fig.~\ref{fig:6ring-shape}. 
Whenever we use a 2-sphere source of large radius $C$, we always find an $S^2\times S^2$
horizon. 
For sufficiently small radius $C$, we do not find the $S^2\times S^2$ horizon, but
we always find an $S^4$ apparent horizon in such a case. 
For middle values of $C$ given by $0.92r_s\leq C\leq 0.96 r_s$, 
 both the $S^4$ and $S^2\times S^2$ apparent horizons are found such that
the $S^4$ horizon encloses $S^2\times S^2$ horizon; see Fig.~\ref{fig:6ring-and-sphere}. 
In order to verify the validity of the inequality, we evaluate 
the following geometric quantities characterizing the size of various submanifolds
in $H$. 
For the $S^4$  horizons, we define 
 three different typical 3-volumes as
\begin{eqnarray}
\!\!\!\!\!\!\!A_s(S_1)\!&=&\!8\pi\!\!\int^{\pi/2}_0 \!\!\psi^2\sqrt{(r_{,\xi})^2\!+\!r^2} ~r^2\cos^2\xi d\xi, \\
\!\!\!\!\!\!\!A_s(S_2)\!&=&\!4\pi^2\!\!\int^{\pi/2}_0 \!\!\psi^2\sqrt{(r_{,\xi})^2\!+\!r^2} ~r^2\sin\xi\cos\xi d\xi,\\
\!\!\!\!\!\!\!A_s(T)\!&=&\!\hbox{max}\left\{8\pi^2 \psi^2 r^3\sin\xi\cos^2\xi;\xi\in\left[0,\frac{\pi}{2}\right]\right\}, 
\end{eqnarray}
where $S_1$ is the 3-sphere section of $H$ on the 4-plane determined by $\eta=0$ and $\eta=\pi$,
 $S_2$ is the 3-sphere section of $H$ on the 4-plane determined by
$\theta=\pi/2$,
and $T$ is the greatest $S^2\times S^1$ on $H$ given by $\xi={\rm const}$.

\begin{figure}[h]
\includegraphics[width=0.8\linewidth]{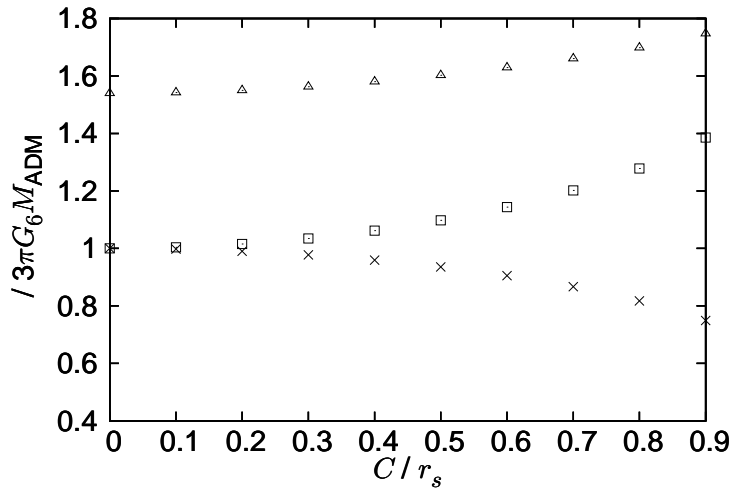}
\caption{$A_s(S_1)$ (squares), $A_s(S_2)$ (crosses), and $A_s(T)$
(triangles) are plotted as a function of radius $C/r_s$. All quantities are normalized by $3 \pi G_6 M_{\rm ADM}$. Although $A_s(S_1)$ and $A_s(T)$ increase with $C$, an $S^4$ black hole with $C\geq 0.97r_s$ does not exist.
\label{fig:6sphere-Area-lastdate}}
\end{figure}

\begin{figure}[h]
\includegraphics[width=0.8\linewidth]{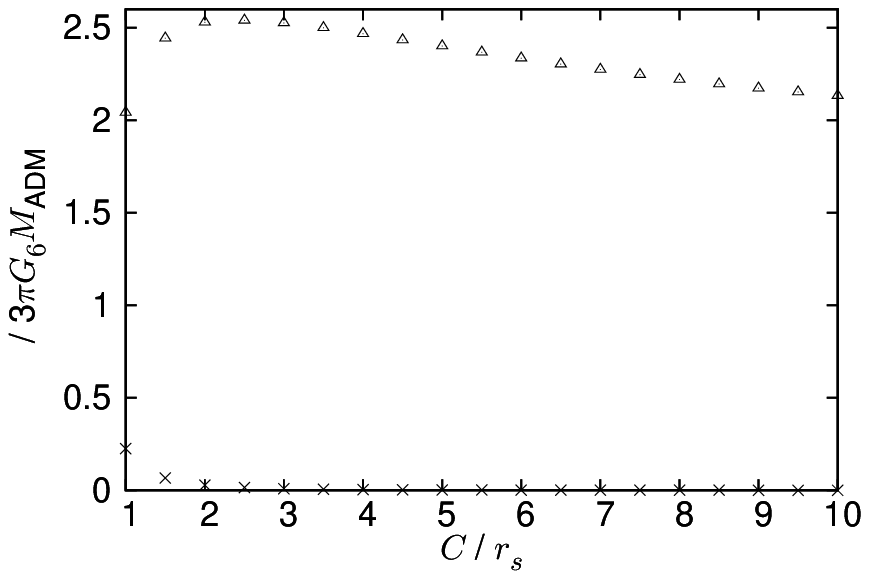}
\caption{$A_n(T_1)$ (crosses) and $A_n(T_2)$ (triangles) 
are plotted as a function of radius $C/r_s$. 
All quantities are normalized by 
$3 \pi G_6 M_{\rm ADM}$.
\label{fig:6ring-Area-latedate}}
\end{figure}

The results of the evaluation of these quantities are shown in 
Fig.~
\ref{fig:6sphere-Area-lastdate}.
For an $S^2\times S^2$ horizon,
we define typical 3-volumes as
\begin{eqnarray}
\!\!\!\!\!\!\!\!\!A_n(T_1)\!\!&=&\!\!2\pi^2 \!\!\int^\pi_0 \!\!\!\psi^2\!\sqrt{(r_{,\xi})^2\!+\!r^2} (r\cos\xi\!+\!C)r\sin\xi d\xi, 
\\
\!\!\!\!\!\!\!\!\!A_n(T_2)\!\!&=&\!\!\!\hbox{max}\!\left\{8\pi^2 \psi^2 (r\cos\xi\!+\!C)^2 r\sin\xi;\xi\!\in\![0,\pi]\right\}\!, 
\end{eqnarray}
where
$T_1$ is the $S^2\times S^1$ section of $H$ obtained by $\theta=\pi/2$,
and $T_2$
is the greatest $S^2\times S^1$ in $H$ among those obtained by $\xi={\rm const}$.
The behavior of these quantities are shown in 
Fig.~
\ref{fig:6ring-Area-latedate}.

From Figs.~\ref{fig:6sphere-Area-lastdate} and \ref{fig:6ring-Area-latedate}, every three-dimensional volume of an apparent horizon in this case satisfies 
\begin{eqnarray}
\frac{(\mbox{3-Volume})}{3\pi G_6M_{\rm ADM}} < O(1),  \label{eq:hoop-con-inequality}
\end{eqnarray}
This shows that the inequality (\ref{eq:hyper-hoop}) 
is satisfied if the mass is regarded as the ADM mass.
We also take into account the possibility that the ``mass" in the hyperhoop conjecture
should be the quasilocal mass $M_{\rm ql}$ instead of $M_{\rm ADM}$.
We define the quasilocal mass as
\begin{eqnarray}
M_{\rm{ql}}=\frac{1}{4\pi}\left(\frac{8\pi^2}{3}\right)^{1/4}V_4^{3/4}.
\end{eqnarray}
where $V_4$ is the area of the apparent horizon, which is given by
\begin{eqnarray}
V_4=8\pi^2\!\!\int^{\pi/2}_0\!\! d\xi \psi^{8/3}\sqrt{(r_{,\xi})^2+r^2} ~r^3\cos^2\xi\sin\xi 
\end{eqnarray}
for an $S^4$ apparent horizon, and by
\begin{eqnarray}
V_4=8\pi^2\!\!\int^\pi_0 \!\!\!d\xi \psi^{8/3}\sqrt{(r_{,\xi})^2\!+\!r^2} (r\cos\xi\!+\!C)^2 r\sin\xi 
\end{eqnarray}
for an $S^2\times S^2$ apparent horizon.

We numerically evaluate the following quantity: 
\begin{eqnarray}
\Gamma=\frac{(\mbox{3-Volume})}{3\pi G_6M_{\rm{ql}}}.
\label{eq:Gamma}
\end{eqnarray}

\begin{figure}[h]
\includegraphics[width=0.8\linewidth]{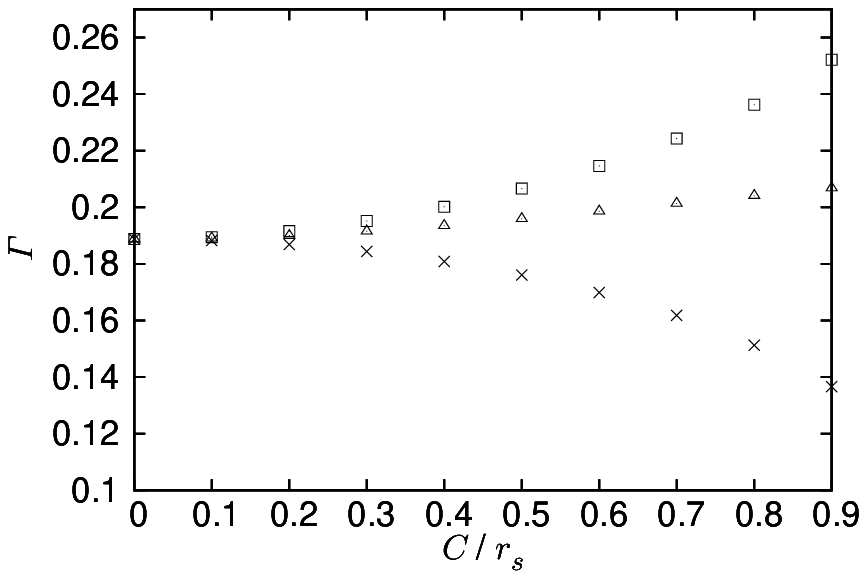}
\caption{$\Gamma$ is plotted as a function of the radius $C/r_s$. $A_s(S_1)$, $A_s(S_2)$, and $A_s(T)$ are depicted by squares, crosses, and triangles, respectively. 
Although $A_s(S_1)$ and $A_s(T)$ increase with $C$, an $S^4$ black hole with $C\geq 0.97r_s$ does not exist.
\label{fig:6S2s-Gamma}}
\end{figure}

\begin{figure}[h]
\includegraphics[width=0.8\linewidth]{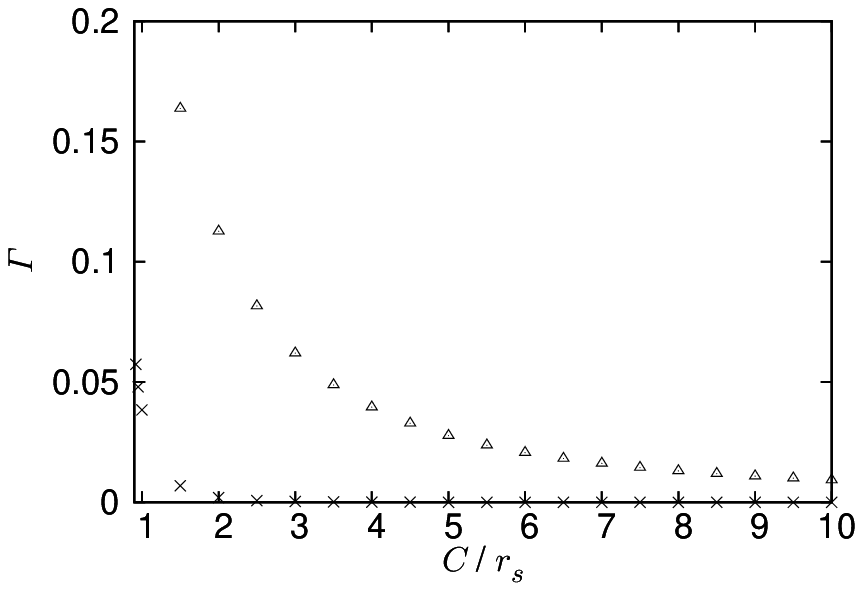}
\caption{$\Gamma$ is plotted as a function of the radius $C/r_s$. $A_n(T_1)$ and $A_n(T_2)$ are depicted by crosses and triangles, respectively.
\label{fig:6S2r-Gamma}}
\end{figure}

We show the results in Figs.~\ref{fig:6S2s-Gamma} and~\ref{fig:6S2r-Gamma}. From Figs.~\ref{fig:6S2s-Gamma} and~\ref{fig:6S2r-Gamma}, for the quasilocal mass both $S^4$ and $S^2\times S^2$ apparent horizon also satisfy 
\begin{eqnarray}
\Gamma < O(1).  \label{eq:quasi-hoop}
\end{eqnarray}
Thus the inequality (\ref{eq:hyper-hoop}) for the quasilocal mass also holds well.

\subsection{$S^1\times S^3$ apparent horizons}
Next, we consider the gravitational field produced by the uniform massive 1-source.
For this purpose, we take the spatial metric as
\begin{eqnarray}
\boldsymbol{g}=\psi^{4/3} \!\left[dx^{2} 
\!+\! x^{2}d\eta^2\!+\!dy^{2}  
\!+\! y^{2}(d\theta^2\!+\!\sin^2 \theta d\phi^2)
\right],  
\end{eqnarray}
which corresponds to the case of $k=1$ in Eq.~(\ref{spatial_metric}).

Let us consider the uniform massive $1$-sphere source of the Euclidean radius $C$ given by
\begin{eqnarray}
\psi^{7/3}\rho=\frac{M_{\rm ADM}\delta(x-C)\delta(y)}{8\pi^2 Cy^2}.
\end{eqnarray}
By solving Eq.~(\ref{Hamiltonian-constraint-1-1}), the conformal factor is obtained in an analytic form as
\begin{eqnarray}
&&\!\!\!\!\!\psi=1\!+\!\frac{3G_6M_{\rm ADM}}{16\pi^2}\!\!\int^{2\pi}_0\!\!\frac{d\theta}{[(x\!-\!C\cos\theta)^2\!+\!y^2\!+\!C^2\sin^2\theta]^{3/2}}\nonumber\\
&&=1+\frac{3G_6M_{\rm ADM}E\left(-\frac{4Cx}{(x-C)^2+y^2} \right)}{4\pi^2[(x+C)^2+y^2]\sqrt{(x-C)^2+y^2}},
\end{eqnarray}
where $E(k)$ is the complete elliptic integral of the second kind,
\begin{eqnarray}
E(k)=\int^{\pi/2}_0\sqrt{1-k\sin^2\theta}d\theta. 
\end{eqnarray}

Now, as is the case in the uniform massive 2-source
, we try to search the location of the apparent horizon in the form (\ref{eq:S2timesS2_ring_embedded}).
Then, the Eq.~ (\ref{location-of-apparenthorizon-k}) for $k=1$ become
\begin{eqnarray}
&&r_{,\xi \xi}-4\frac{(r_{,\xi})^{2}}{r}-3r-\frac{(r_{,\xi})^{2} +r^{2}}{r} \nonumber \\
&&\hspace{0.2cm} \times\left( \frac{r_{,\xi}\sin \xi \!+\! r\cos \xi}{r\cos \xi \!+\!C }\!-\!2\frac{r_{,\xi}}{r}\cot \xi \right)\!-\!\frac{8}{3}\frac{(r_{,\xi})^{2} \!+\!r^{2}}{r} \nonumber \\
&&\hspace{0.2cm}\times\left\{(r_{,\xi}\sin \xi \!+\! r\cos \xi )\frac{\psi_{,x}}{\psi}\!-\!(r_{,\xi}\cos \xi \!-\! r\sin \xi )\frac{\psi_{,y}}{\psi}\right\} \nonumber\\
&&\hspace{0.1cm}=0.
\label{eq:S1-the-location-of-apparent-horizon}
\end{eqnarray}
The location of the $S^1\times S^3$ apparent horizon of a black hole is obtained by solving the Eq.~(\ref{eq:S1-the-location-of-apparent-horizon}) 
for $r(\xi)$ with the boundary condition $r_{,\xi}=0$ for $\xi=0,\pi$. 

We also search for the $S^4$ apparent horizon in the form (\ref{eq:S2timesS2_sphere_embedded}).
Then we solve the minimal surface equation
\begin{eqnarray}
&&r_{,\xi \xi}-4\frac{(r_{,\xi})^{2}}{r}-3r-\frac{(r_{,\xi})^{2} +r^{2}}{r} \nonumber \\
&&\hspace{0.2cm} \times\left( \frac{r_{,\xi}\sin \xi \!+\! r\cos \xi}{r\cos \xi  }\!-\!2\frac{r_{,\xi}}{r}\cot \xi \right)\!-\!\frac{8}{3}\frac{(r_{,\xi})^{2} \!+\!r^{2}}{r} \nonumber \\
&&\hspace{0.2cm}\times\left\{(r_{,\xi}\sin \xi \!+\! r\cos \xi )\frac{\psi_{,x}}{\psi}\!-\!(r_{,\xi}\cos \xi \!-\! r\sin \xi )\frac{\psi_{,y}}{\psi}\right\} \nonumber\\
&&\hspace{0.1cm}=0,
\label{eq:S1-the-location-of-apparent-horizon2}
\end{eqnarray}
with the boundary condition $r_{,\xi}=0$ for $\xi=0, \pi/2$.

\begin{figure}[h]
\begin{center}
\includegraphics[width=0.8\linewidth]{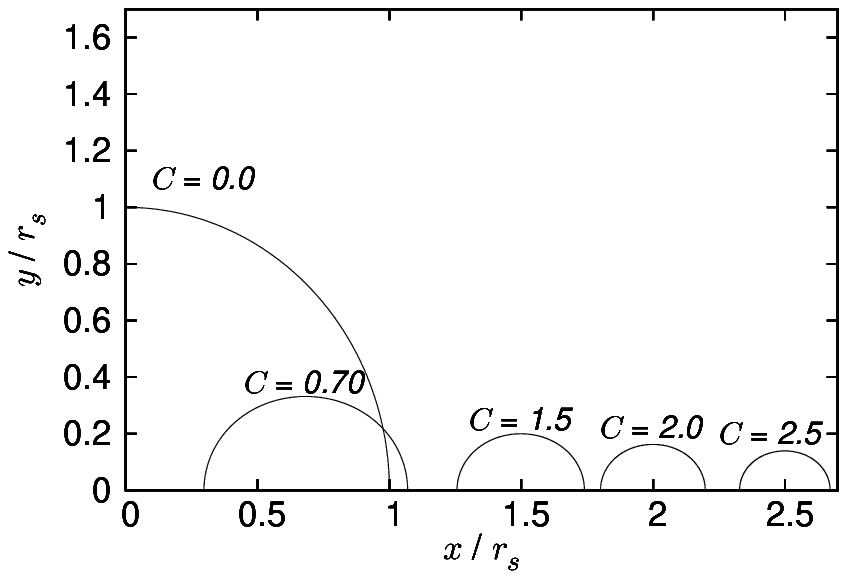}
\caption{$S^1\times S^3$ apparent horizons for $C=0.7r_s$, $1.5r_s$, $2.0r_s$, $2.5r_s$ and the spherically symmetric black hole for a point source at the origin $x=0.0r_s$ are depicted in $(x,y)$-plane.
 The $S^1\times S^3$ apparent horizon for $C=0.67r_s$ is not found.\label{fig:6S1r-shape}} 
 \end{center}
\end{figure}

\begin{figure}[h]
\begin{center}
\includegraphics[width=0.8\linewidth]{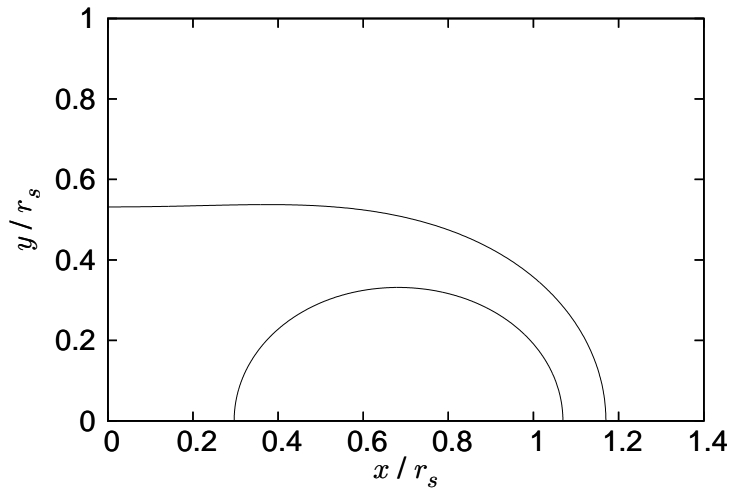}
\caption{We can find both the $S^4$ and $S^1\times S^3$ apparent horizon for $0.68r_s\leq C\leq 0.72r_s$. Both horizons for $C=0.70r_s$ are depicted in the $(x,y)$ plane. $S^1\times S^3$ apparent horizon for $C=0.73r_s$ is not found.\label{fig:6S1-sphere-and-ring}}
\end{center}
\end{figure}

We show the shape of the $S^1\times S^3$ apparent horizon by solving Eq.~(\ref{eq:S1-the-location-of-apparent-horizon}) and Eq.~(\ref{eq:S1-the-location-of-apparent-horizon2}) with Fig.~ \ref{fig:6S1r-shape}. When we put 
a 1-sphere source of large radius, we always find the $S^1\times S^3$ apparent horizon. For small radius, we find the $S^4$ black hole. 
We also found that there exists both the $S^4$ and $S^1\times S^3$ apparent horizon for $0.68r_s\leq C\leq 0.72r_s$ in the 
Fig.~\ref{fig:6S1-sphere-and-ring}.

We evaluate typical area scales for the case of the $S^4$ apparent horizon,
\begin{eqnarray}
\!\!\!\!A_s(S_1)\!\!\!&=&\!\!8\pi \!\!\int^{\pi/2}_0 \!\!\psi^2\sqrt{(r_{,\xi})^2+r^2} ~r^2\sin^2\xi d\xi,  \\
\!\!\!\!A_s(S_2)\!\!\!&=&\!\!2\pi^2 \!\!\int^{\pi/2}_0 \!\!\psi^2\sqrt{(r_{,\xi})^2+r^2} ~r^2\sin2\xi d\xi,  \\
\!\!\!\!A_s(T)\!\!&=&\!\!\hbox{max}\left\{8\pi^2 \psi^2 r^3\cos\xi\sin^2\xi;\xi\in\left[0,\frac{\pi}{2}\right]\right\}\!,  
\end{eqnarray}
where $S_1$ is the 3-sphere section of $H$ on the 4-plane determined by $\eta=0$ and $\eta=\pi$,
$S_2$ is the 3-sphere section of $H$ on the 4-plane determined by $\theta=\pi/2$,
and $T$ is the greatest $S^1 \times S^2$ on $H$ given by $\xi= \mbox{const}$.

\begin{figure}[h]
\begin{center}
\includegraphics[width=0.8\linewidth]{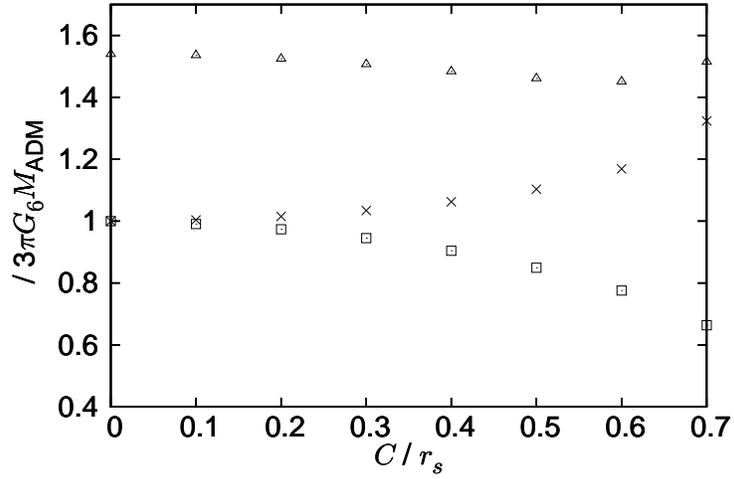}
\caption{$A_s(S_1)$(squares), $A_s(S_2)$(crosses), and $A_s(T)$
(triangles) are plotted as a function of radius $C/r_s$.
All quantities are normalized by 
$3 \pi G_6 M_{\rm ADM}$. Although $A_s(S_2)$ and $A_s(T)$ increase as a function of $C$, a $S^4$ 
black hole with $C\geq 0.73r_s$ does not exist.\label{fig:6S1s-Area}}
\end{center}
\end{figure}

\begin{figure}[h]
\begin{center}
\includegraphics[width=0.8\linewidth]{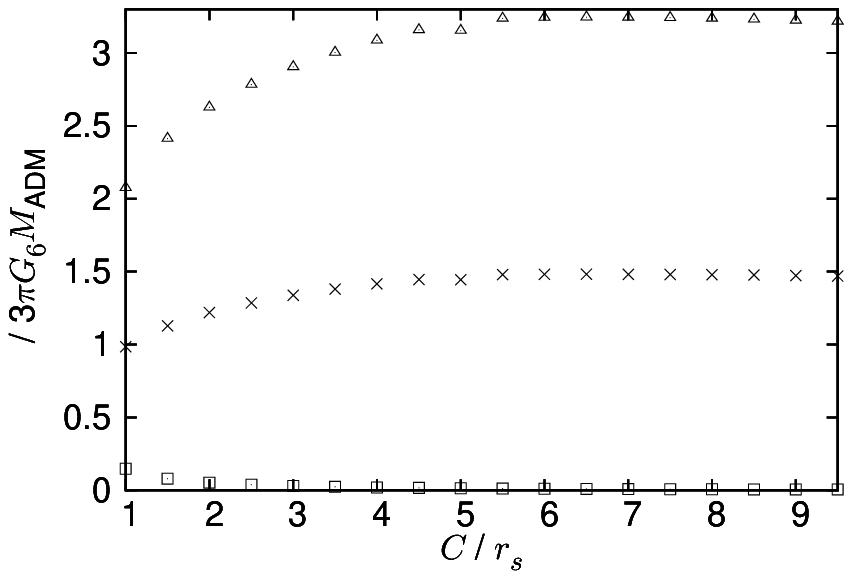}
\caption{$A_n(S)$(squares), $A_n(T_1)$(crosses), and $A_n(T_2)$ (triangles)   
are plotted as a function of radius $C/r_s$.
All quantities are normalized by 
$3 \pi G_6 M_{\rm ADM}$.\label{fig:6S1r-Area-and-Vol2}}
\end{center}
\end{figure}

\begin{figure}[h]
\begin{center}
\includegraphics[width=0.8\linewidth]{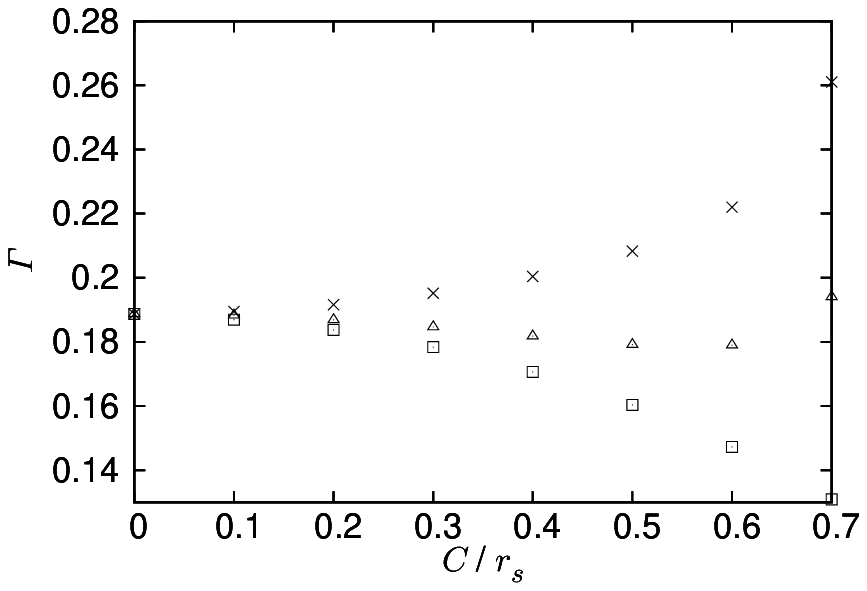}
\caption{$\Gamma$ is plotted as a function of the radius $C/r_s$. $A_s(S_1)$, $A_s(S_2)$, and $A_s(T)$ are depicted by squares, crosses, and triangles, respectively. 
Although $A_s(S_2)$ and $A_s(T)$ increase as a function of $C$, an $S^4$ 
black hole with $C\geq 0.73r_s$ does not exist.\label{fig:6S1s-Gamma}}

\end{center}
\end{figure}

\begin{figure}[h]
\begin{center}
\includegraphics[width=0.8\linewidth]{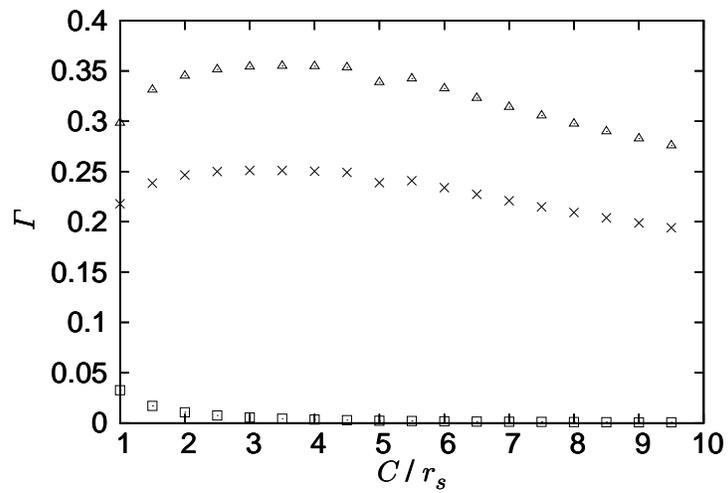}
\caption{$\Gamma$ is plotted as a function of the radius $C/r_s$. $A_s(S)$, $A_s(T_1)$, and $A_s(T_2)$ are depicted by squares, crosses, and triangles, respectively. 
\label{fig:6S1r-Gamma}}
\end{center}
\end{figure}

For the case of the nontopological apparent horizon ($\simeq S^1\times S^3$), we evaluate typical area scales,

\begin{eqnarray}
\!\!\!\!\!\!\!\!\!\!\!\!A_n(S)\!\!\!&=&\!\!\!4\pi \!\!\int^\pi_0 \!\!\psi^2\!\sqrt{(r_{,\xi})^2\!+\!r^2} r^2\sin^2\xi d\xi,\\
\!\!\!\!\!\!\!\!\!\!\!\!A_n(T_1)\!\!\!&=&\!\!\!2\pi^2 \!\!\int^\pi_0 \!\!\psi^2\!\sqrt{(r_{,\xi})^2\!+\!r^2}(r\cos\xi\!+\!C)r\sin\xi d\xi, \\
\!\!\!\!\!\!\!\!\!\!\!\!A_n(T_2)\!\!\!&=&\!\!\hbox{max}\!\left\{8\pi^2 \psi^2 (r\cos\xi\!+\!C)r^2\sin^2\xi;\xi\!\in\![0,\pi]\right\}\!\!, 
\end{eqnarray}

where $S$ is the 3-sphere section of $H$ on the 4-plane determined by $\eta=0$ and $\eta=\pi$,
$T_1$ is the $S^2 \times S^1$ section of $H$ obtained by $\theta=\pi/2$,
and $T_2$ is the greatest $S^2\times S^1$ on $H$ given by $\xi={\rm const}$.

We evaluate the above quantities as functions of the source radius $C$ in Figs.~\ref{fig:6S1s-Area} and \ref{fig:6S1r-Area-and-Vol2}. As in the previous section, the inequality (\ref{eq:hoop-con-inequality}) for every 3-volume is satisfied 
from Figs.~\ref{fig:6S1s-Area} and \ref{fig:6S1r-Area-and-Vol2}. 
Thus, inequality (\ref{eq:hyper-hoop}) for the ADM mass is satisfied.

We evaluate $\Gamma$ in Eq.~(\ref{eq:Gamma}), where $V_4$ is given by
\begin{eqnarray}
V_4=8\pi^2\!\int^{\pi/2}_0 \!\!d\xi \psi^{8/3}\sqrt{(r_{,\xi})^2+r^2} ~r^3\cos\xi\sin^2\xi 
\end{eqnarray}
for an $S^4$ apparent horizon, and
\begin{eqnarray}
\!\!V_4=8\pi^2\!\!\int^\pi_0 \!\!d\xi \psi^{8/3}\!\sqrt{(r_{,\xi})^2\!+\!r^2}(r\cos\xi\!+\!C)(r\sin\xi)^2 
\end{eqnarray}
for an $S^1\times S^3$ apparent horizon.
We show the results of the evaluation of $\Gamma$ in Figs.~\ref{fig:6S1s-Gamma} and~\ref{fig:6S1r-Gamma}. As in the $S^2\times S^2$ case, inequality (\ref{eq:hyper-hoop}) for the quasilocal mass is satisfied, which can be seen from Figs.~\ref{fig:6S1s-Gamma} and~\ref{fig:6S1r-Gamma}.

\section{Conclusion}
We have investigated the validity of the inequality 
\begin{eqnarray*}
V_3\lesssim G_6 M
\end{eqnarray*}
in the six-dimensional black hole space-times,
in terms of  the momentarily static, conformally flat initial data set 
 on the five-dimensional Cauchy surface.

When we put the uniform massive 2-sphere source on the Cauchy surface, 
 we find the black hole is homeomorphic to $S^2\times S^2$
for the source of sufficiently large radius. 
On the other hand, 
for the source of small radius, we found the black holes are homeomorphic to $S^4$. 
It is also verified that both the $S^4$ and $S^2\times S^2$ horizons
are found for the  intermediate parameter values.
We also consider the gravitational field produced by a uniform massive circle source.
We find  the black hole homeomorphic to  $S^1\times S^3$ for large source 
radius and the spherical black holes homeomorphic to
 $S^4$ for small source radius. 
Again, there is an intermediate parameter range where both
 $S^4$ and $S^1\times S^3$ horizons coexist.
 Here, existence of an $S^1 \times S^3$ apparent horizon can be 
intuitively understood as follows.
 We can expect the existence of an apparent horizon for
a large circle source,
since the geometry around such a large circle would be well approximated by
that of a straight black string.
Accordingly, this type of apparent horizon should be homeomorphic with $S^1 
\times S^3$.
Similarly the existence of an $S^2\times S^2$ horizon for a large $2$-sphere source
 is plausible due to the existence of the black brane solutions. For all cases above, we evaluated the size of
the characteristic 3-volume of apparent horizons.
We cannot find any evidence that the hyperhoop conjecture (\ref{eq:hyper-hoop}) 
is false.

\section*{acknowledgments}
The authors are grateful to Daisuke Ida for his comments and encouragement.


\begin{thebibliography}{99}
\bibitem{Thorne}
K. S.~Thorne, 
in {\it Magic Without Magic},
ed. J.R.~Klauder (Freeman, San. Francisco, 1972).

\bibitem{Ida-Nakao}
D.~Ida and K.~Nakao, 
Phys. Rev. D 66, 064026 (2002).

\bibitem{Barrabes}
C.~Barrab$\grave{e}$s, V.P.~Frolov and E.~Lesigne, 
Phys. Rev. D 69, 101501 (2004) .

\bibitem{Yamada-Shinkai}
Y.~Yamada and H.~Shinkai, 
Class. Quant. Grav. 27, 045012 (2010).

\bibitem{Chul-Nakao-Ida}
C-M.~Yoo, K.~Nakao and D.~Ida, 
Phys. Rev. D 71, 104014 (2005).

\bibitem{Yoshino-Numbu}
H.~Yoshino and Y.~Nambu, 
Phys. Rev. D67 024009 (2003).

\bibitem{Cvetic}
M.~Cvetic, G.W.~Gibbons,~C.N.~Pope,
Class. Quant. Grav. 28, 195001 (2011).

\bibitem{Cai-Galloway}
M.~Cai and G.J.~Galloway,
Class. Quant. Grav. 18, 2707 (2001).

\bibitem{Haw}
S.W.~Hawking,
Commun. Math. Phys. 25, 152 (1972).

\bibitem{Galloway-Schoen}
G.J.~Galloway and R.~Schoen,
Commun. Math. Phys. 266, 571 (2006).

\bibitem{Emp}
R.~Emparan and H.~Reall,
Phys. Rev. Letters 88, 101101 (2002).

\end{thebibliography}
\end{document}